\documentclass[a4paper,twocolumn,aps,prl,showpacs]{revtex4-1}
\usepackage{bbm}
\usepackage{amsmath}
\usepackage{verbatim}
\usepackage{graphicx}
\usepackage{times}
\usepackage{amssymb}
\usepackage{epsfig}
\usepackage{graphicx}
\usepackage{bm}
\usepackage{times}
\usepackage{txfonts}
\usepackage{dsfont}

\begin{document}

\title{Atom-mirror cooling and entanglement using cavity Electromagnetically
Induced Transparency}
\author{Claudiu Genes$^1$, Helmut Ritsch$^1$, Michael Drewsen$^2$ and Aur\'elien Dantan$^2$}
\date{\today }
\affiliation{$^1$Institute for Theoretical Physics, University of
Innsbruck, Technikerstrasse 25, A-6020 Innsbruck, Austria\\
$^2$QUANTOP, Danish National Research Foundation Center for Quantum
Optics, Department of Physics and Astronomy, University of Aarhus,
DK-8000 Aarhus C, Denmark}

\begin{abstract}
We investigate a hybrid optomechanical system comprised of a mechanical oscillator and an atomic
3-level ensemble within an optical cavity. We show that a suitably tailored cavity field
response via Electromagnetically Induced Transparency (EIT) in the atomic medium allows for strong
coupling of the mechanical mirror oscillations to the collective atomic ground-state spin. This
facilitates ground-state cooling of the mirror motion, quantum state mapping and robust atom-mirror
entanglement even for cavity widths larger than the mechanical oscillator frequency.
\end{abstract}

\maketitle

The past years have witnessed tremendous progress towards the control of
mechanical motion at the quantum limit in micro- and nano-optomechanical
systems~\cite{opto}. While cavity optomechanical phenomena
are traditionally investigated with solid-state optomechanical systems -
micromirrors, cantilever tips, toroidal resonators, movable membranes, etc.
- cold atomic gasses placed in high-finesse optical cavities~\cite{atomopto} have also successfully been used to implement
equivalent Hamiltonians at ultralow temperatures. Consequently, several
proposals suggested a combination of both approaches to realize hybrid
optomechanical systems~\cite{hybrid,genes09,hammerer10}
, in which well-controlled atomic systems can be interfaced with solid-state
mechanical resonators. These can benefit from the well-established atomic
physics toolbox for cooling, trapping, state preparation, control and
readout and allow to properly tailor the atom-cavity response function.

We propose here a hybrid system composed of a mechanical oscillator and a 3-level atomic medium
operated in an Electromagnetically Induced Transparency (EIT) configuration within the optical
cavity~\cite{lukin98}. We show how the cavity field response can be tailored~\cite{genes09,elste09}
by the EIT interaction with the medium in order to strongly couple the motion of the mechanical
oscillator to the collective atomic ground-state spin. The sharp and tunable nature of the cavity
field EIT resonance allows for efficiently addressing either the Stokes or anti-Stokes motional
sidebands of the movable mirror (which is reminiscent of EIT cooling of ions~\cite{morigi00}), even
in the bad-cavity limit, i.e. when its mechanical resonance frequency is much smaller than the
cavity linewidth. We show in particular how to engineer ''beamsplitter"- or ''down-conversion"-type
Hamiltonians~\cite{nielsen00} between the movable mirror motion and the collective atomic spin,
which can be exploited for efficient optomechanical cooling, quantum state mapping or robust
atom-mirror entanglement generation. Such interactions would be especially appealing for
low-mechanical resonance frequency (sub-MHz) oscillators, such as movable
membranes~\cite{membrane}, coupled to cold atoms/Bose-Einstein
condensates~\cite{atomopto,coldatomcavity,hammerer10} or ion crystals~\cite{herskind09} in
low-finesse optical cavities.

\begin{figure}[t]
\includegraphics[width=0.95\columnwidth]{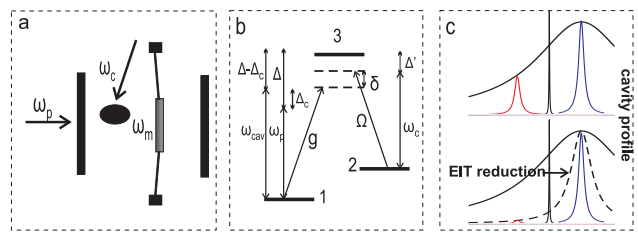}
\caption{(a) Hybrid optomechanical system composed of an atomic ensemble
and a mechanical oscillator enclosed in an optical cavity. The cavity field is
coupled to the mechanical oscillator motion via radiation pressure and to an atomic
transition, while an external control field allows for achieving EIT in the
atomic medium. (b) Internal atomic level structure. (c) Cavity field
transmission frequency profile for (un)resolved sideband cooling of the
mirror motion in the bad cavity limit (upper) and cavity EIT-resolved
sideband cooling (lower).}
\label{scheme}
\end{figure}

\textit{Model} Let us consider an ensemble of 3-level atoms/ions in a $%
\Lambda$ configuration coupled to a control laser and a cavity field mode on
the two upwards transitions. Via the cavity mode the atoms interact with
a movable membrane within the cavity or in case of a single-ended cavity
with one movable end-mirror. The atomic operators are denoted by $\sigma_{\alpha\beta}^{(j)}$ ($j=1-N$). The level
frequency separations are $\omega _{13}$, $\omega _{23}$ as optical
transitions and $\omega _{12}$ in the microwave domain. The cavity field $a$
is driven at $\omega _{p}$, close to a cavity resonance $\omega _{cav}$. A
membrane/mirror vibrational mode at frequency $\omega _{m}$ and with ladder
operators $b$,$b^{\dagger }$ can be excited by the radiation pressure of the
cavity field. The free Hamiltonian is (with $\hbar=1$) $\mathcal{H}_{0}=\omega
_{21}\sum_{j}\sigma _{22}^{(j)}+\omega _{31}\sum_{j}\sigma
_{33}^{(j)}+\omega _{cav}a^{\dagger }a+\omega _{m}b^{\dagger }b$. With an
extra control laser driving on the $2-3 $ transition at frequency $\omega
_{c}$, the atom-field interaction is $\mathcal{H}_{at-f}=-g(\sum_{j}\sigma
_{31}^{(j)}a+\mathrm{h.c.})-\Omega (\sum_{j}\sigma _{32}^{(j)}e^{-i\omega
_{c}t}+\mathrm{h.c.}),$ where $g$ is the single atom-cavity field coupling strength and $\Omega $ the control field Rabi frequency. The
optomechanical interaction part contains the bare optomechanical coupling $%
G_{0}$, $\mathcal{H}_{f-m}=-G_{0}a^{\dagger }a\left( b^{\dagger }+b\right) .$

We consider a typical EIT regime for which the cavity field is much weaker
than the control field ($g\left\vert \left\langle a\right\rangle \right\vert
\ll \Omega $), and most of the atoms are in level 1. This allows us to make
the standard bosonization approximation and map the spin algebra to a
harmonic oscillator algebra via the transformation $1/\sqrt{N}\sum_{j}\sigma
_{12,13}^{(j)}\rightarrow c_{2,3}$ with $[c_{2,3},c_{2,3}^{\dagger }]=1$. In a rotating frame that redefines dynamics in terms of
detunings: $\Delta _{cav}=\omega _{cav}-\omega _{p}$, $\Delta =\omega
_{31}-\omega _{p}$, $\Delta ^{\prime }=$ $\omega _{32}-\omega _{c}$ and $%
\delta =\Delta -\Delta ^{\prime }$, one can derive the following set of
coupled equations of motion that will be the starting point of our
calculations
\begin{subequations}
\begin{eqnarray}
\dot{c}_{3} &=&-\left( \gamma +i\Delta \right) c_{3}+ig_{N}a+i\Omega
c_{2}+c_{3}^{in},  \label{eq:c_3} \\
\dot{c}_{2} &=&-\left( \gamma _{c}+i\delta \right) c_{2}+i\Omega
c_{3}+c_{2}^{in}, \\
\dot{a} &=&-\left( \kappa +i\Delta _{cav}\right) a+ig_{N}c_{3}+iG_{0}a\left(
b^{\dagger }+b\right) +a_{in},  \label{eq:a} \\
\dot{b} &=&-\left( \gamma _{m}+i\omega _{m}\right) b+\gamma _{m}b^{\dagger
}+G_{0}a^{\dagger }a+b_{in},  \label{eq:b}
\end{eqnarray}
\end{subequations}
where $\kappa $, $2\gamma _{m}$, $\gamma $ and $\gamma _{c}$ are the decay
rates of the cavity field, the mirror, and the dipoles on the 3-1 and 2-1 transitions,
respectively. Zero-mean valued Langevin noise terms have been added to the equations of
motion with the following relevant correlation functions $\langle
c_{3}^{in}\left( t\right) c_{3}^{in,\dagger }\left( t^{\prime }\right)
\rangle =\gamma \delta (t-t^{\prime })$, $\langle c_{2}^{in}\left( t\right)
c_{2}^{in,\dagger }\left( t^{\prime }\right) \rangle =\gamma _{c}\delta
(t-t^{\prime })$, $\langle a_{in}\left( t\right) a_{in}^{\dagger }\left(
t^{\prime }\right) \rangle =\kappa \delta (t-t^{\prime })$, $\langle
b^{in}\left( t\right) b^{in,\dagger }\left( t^{\prime }\right) \rangle
=2\gamma _{m}(n_{i}+1)\delta (t-t^{\prime })$ and $\langle b^{in,\dagger
}\left( t\right) b^{in}\left( t^{\prime }\right) \rangle =2\gamma
_{m}n_{i}\delta (t-t^{\prime })$. The occupancy of the mechanical
resonator imposed by the external thermal reservoir is denoted here by $n_{i}$.

\textit{Dressed cavity field response} In steady state the intracavity field
mean value is given by $\left\langle a\right\rangle =\langle a_{in}\rangle
/\left( \kappa +i\Delta _{c}-i\chi _{EIT}\right) $ where $\Delta _{c}=\Delta
_{cav}-G^{2}/\omega _{m}$ ($G=G_{0}\langle a\rangle $) and the EIT medium
susceptibility is
\begin{equation}
\chi _{EIT}=\frac{ig_{N}^{2}}{\gamma +i\Delta +\frac{\Omega ^{2}}{\gamma
_{c}+i\delta }}.
\end{equation}%
For a strongly absorbing medium ($g_{N}>\kappa ,\gamma $) the cavity will only be transparent in a
narrow frequency range around the two-photon (EIT) resonance. This emulates a cavity substantially
narrower than its natural linewidth $2\kappa$. Under the assumptions $\Omega ^{2}\gg
\gamma_{c}\gamma $ and $\Omega \gg \gamma _{c}$ the cavity transmission spectrum becomes a
Lorentzian peak centered around $\delta =0$ with a modified halfwidth
\begin{equation}
\kappa _{EIT}\simeq \gamma _{c}+\kappa \frac{\Omega ^{2}}{g_{N}^{2}}
\end{equation}
An effective sharpening of the cavity response around the two-photon atomic
resonance can thus be obtained if narrow atomic resonances ($\gamma _{c}\ll
\kappa $) and strong atom-cavity coupling strengths ($g_{N}\gg \Omega $) are
used~\cite{lukin98}. This tailoring of the cavity field response will now be exploited for
engineering coupling between the atoms and the mirror motion. For a simple
physical understanding one can Fourier analyze Eqs.~(\ref{eq:a}) to derive the
cavity response in the frequency domain in the presence of atoms. The result is
plotted in Fig.~\ref{scheme}c where, for example, the EIT sharpening of the cavity profile
around the blue sideband shows the mechanism through which the inhibition of
the red sideband improves cooling as compared to the situation where no
atoms are present.

Eqs.~(\ref{eq:c_3}-\ref{eq:b}) can be linearized around their steady state
mean values and the variance matrix of the quantum fluctuations of all
observables can be calculated numerically~\cite{gardiner}. The most interesting physical situations correspond to tuning the dressed
cavity field resonances to either the anti-Stokes or the Stokes motional
sidebands. Here, the analysis is most conveniently performed by moving to the
corresponding rotating frames.

\textit{Anti-Stokes sideband resonance: cooling and state mapping} We first
assume that the cavity and the atomic two-photon detunings are matched to
the anti-Stokes motional frequency, $\delta =\Delta _{c}=\omega _{m}$. Hence
emission of a cavity photon amounts to absorbing a mirror vibrational
quantum. In the frame rotating at $\omega _{m}$ and neglecting off-resonant
interactions the equations for the fluctuations read
\begin{subequations}
\label{RWA}
\begin{eqnarray}
\dot{\tilde{c}}_{3} &=&-\gamma \tilde{c}_{3}+ig_{N}a+i\Omega \tilde{c}_{2}+%
\tilde{c}_{3}^{in}, \\
\dot{\tilde{c}}_{2} &=&-\gamma _{c}\tilde{c}_{2}+i\Omega \tilde{c}_{3}+%
\tilde{c}_{2}^{in}, \\
\dot{\tilde{a}} &=&-\kappa \tilde{a}+ig_{N}\tilde{c}_{3}+iG\tilde{b}+\tilde{a%
}_{in}, \\
\dot{\tilde{b}} &=&-\gamma _{m}\tilde{b}+iG\tilde{a}+\tilde{b}_{in}.
\end{eqnarray}
\end{subequations}
where $\tilde{o}=oe^{-i\omega _{m}t}$. We look at the effective interaction between $\tilde{c}_{2}$
and $\tilde{b}$ in the regime when $\gamma ,\kappa \gg \gamma _{c},\gamma _{m},\omega _{m}$, i.e.
such that $\tilde{c}_{3}$ and $\tilde{a}$ are the fast variables that can be adiabatically
eliminated. We first identify two rates that play an important role in the process as the optical
cooling rate $\Gamma _{O}=G^{2}/\kappa $, and the excited-ground state decay rate $\Gamma
_{E}=\Omega ^{2}/\gamma $, with corresponding normalized rates\ $\gamma _{O}=\Gamma _{O}/(1+C)$ and
$\gamma _{E}=\Gamma _{E}/(1+C)$, where $C=g_N^2/\kappa\gamma$ is the cooperativity parameter. We
can now write for the reduced bipartite system
\begin{subequations}
\begin{eqnarray}
\dot{\tilde{c}}_{2} &=&-\left( \gamma _{c}+\gamma _{E}\right) \tilde{c}_{2}-i%
\sqrt{C\gamma _{E}\gamma _{O}}\tilde{b}+\bar{c}_{2}^{in}, \\
\dot{\tilde{b}} &=&-\left( \gamma _{m}+\gamma _{O}\right) \tilde{b}-i\sqrt{%
C\gamma _{E}\gamma _{O}}\tilde{c}_{2}+\bar{b}_{in},
\end{eqnarray}
\end{subequations}
which show the renormalized bare effective decay rates of the system $%
\gamma _{c}+\gamma _{E}$ and $\gamma _{m}+\gamma _{O}$ together with the
coupling rate $\sqrt{C\gamma _{E}\gamma _{O}}$. The effective Langevin noise terms contain
contributions from all the noise processes in the system and are expressed
as $\bar{c}_{2}^{in}=-i\sqrt{{\gamma _{E}}/({1+C)}}c_{3}^{in}/\sqrt{\gamma }-%
\sqrt{{\gamma _{E}C}/({1+C)}}{a_{in}}/\sqrt{\kappa }+c_{2}^{in},\bar{b}%
_{in}=i\sqrt{\gamma _{O}/({1+C)}}a_{in}/\sqrt{\kappa }-\sqrt{\gamma _{O}C/({%
1+C)}}c_{3}^{in}/\sqrt{\gamma }+b_{in}.$ The only non-vanishing correlations
are $\langle \bar{c}_{2}^{in}\left( t\right) \bar{c}_{2}^{in,\dagger }\left(
t^{\prime }\right) \rangle =2(\gamma _{E}+\gamma _{c})\delta (t-t^{\prime })$%
, $\langle \bar{b}^{in,\dagger }\left( t\right) \bar{b}^{in}\left( t^{\prime
}\right) \rangle =2\gamma _{m}n_{i}\delta (t-t^{\prime })$ and $\langle \bar{%
b}^{in}\left( t\right) \bar{b}^{in,\dagger }\left( t^{\prime }\right)
\rangle =2[\gamma _{O}+\gamma _{m}(n_{i}+1)]\delta (t-t^{\prime }).$ An
effective Hamiltonian for the atomic ground state coherence-mechanical
motion can then be expressed as
\begin{equation}
H_{AS}\simeq \sqrt{C\gamma _{E}\gamma _{O}}\left( \tilde{b}^{\dagger }\tilde{%
c}_{2}+\tilde{b}\tilde{c}_{2}^{\dagger }\right) ,
\end{equation}%
and takes the form of the beamsplitter-like interaction extensively used
in quantum optics and quantum information. The investigation of the
different timescales in the system leads to the identification of two
regimes: \textit{i) a cooling regime}, for $\gamma _{O}$ $\ll \gamma _{E}$%
, and \textit{ii) a state transfer (strong coupling) regime}, for $\sqrt{%
C\gamma _{E}\gamma _{O}}\gg \gamma _{E},\gamma _{O},\gamma _{c},\gamma
_{m}n_{i}$, which we analyze analytically and numerically in the following.

\begin{figure}[t]
\includegraphics[width=0.95\columnwidth]{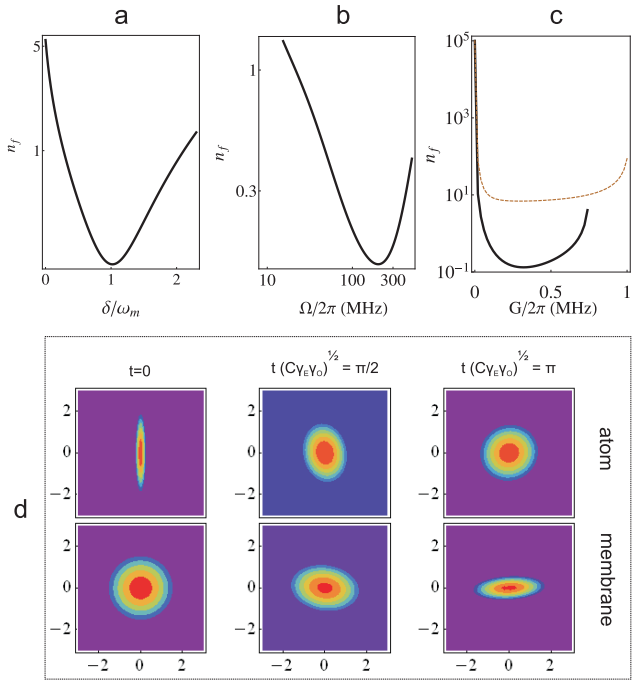}
\caption{\textit{Cavity EIT cooling:} (a) Logarithmic plot of the final occupancy number $n_{f}$ as a
function of normalized two-photon detuning $\delta/\omega_{m}$. (b) Variation of $n_f$ with
$\Omega$ for $\protect\delta=\protect\omega_m$. (c) Variation of $n_f$ with $%
G$ for standard self-cooling with no atoms ($\Delta_{c}=\protect\kappa /2$,
dashed line) and cavity EIT cooling ($\protect\delta=\protect\omega_m$, $%
\Omega=(2\protect\pi) 300$ MHz, solid line). \textit{State mapping}: (d) Time evolution of the atomic
and mechanical oscillator Wigner functions for an initially squeezed ground-state atomic spin. See text for parameters.}
\label{cooling}
\end{figure}

When $\kappa _{EIT}\gg \gamma _{c},\Gamma _{O}$ one can treat the
atom-cavity subsystem as an effective bath for the mechanical degree of
freedom~\cite{selfcooling}. The sharpening of
the cavity response (EIT window) can inhibit Stokes scattering leading to
resolved sideband cooling. Assuming a bad cavity $\kappa \gg \omega _{m}$
(for which direct cavity-induced optomechanical cooling would be slow and
inefficient) we first assume that $\kappa _{EIT}\ll \omega _{m}$ to resolve
sidebands. To avoid entering the strong coupling regime where the blue
sideband spectrum gets wider than the EIT peak we also assume that $\kappa
_{EIT}>\sqrt{C\gamma _{E}\gamma _{O}}$. Under such conditions the cooling
rate is, to a good approximation, $\Gamma _{O}(\gamma _{E}/\kappa _{EIT})$,
while the mirror heating rate is $2\gamma _{m}n_{i}+\gamma _{O}$. We can now
write a simple expression for the final mechanical occupancy
\begin{equation}
n_{f}\simeq \frac{\gamma _{m}}{\gamma _{m}+\Gamma _{O}}n_{i}+\frac{\gamma
_{c}}{2\kappa _{EIT}},  \label{eq:n_f}
\end{equation}%
which can be well-below unity for sufficiently strong optomechanical
coupling and narrow cavity EIT resonances. Note that the
ultimate cooling limit in $\gamma _{c}/2\kappa _{EIT}$ is however bounded by
the resolved sideband condition $\kappa _{EIT}\ll \omega _{m}$.

In the regime where the coherent coupling rate $\sqrt{C\gamma _{E}\gamma _{O}}
$ becomes larger than the effective decay rates, coherent state transfer is
possible between the mirror and the ground state coherence. The conditions
for achieving this goal can be summarized by the following double inequality
\begin{equation}
\frac{\kappa }{g_{N}}\ll \frac{G}{\Omega }\ll \frac{g_{N}}{\gamma }.
\label{eq:strongcouplinginequality}
\end{equation}%
To illustrate the cooling and state transfer regime more clearly we focus now on a numerical
example. We take $N=10^{8}$ Rb atoms ($\gamma =(2\pi )3$ MHz) coupled to an optical cavity with
$\kappa =(2\pi )1$ MHz and $g=(2\pi )100$ kHz and assume a ground state decoherence rate of $\gamma
_{c}\sim (2\pi )1$ kHz, as obtained e.g. with ion crystals~\cite{albert11}. We assume a mechanical
resonator with oscillation frequency $\omega _{m}=(2\pi )200$ kHz and mechanical quality factor
$Q_{m}\sim 10^{7}$~\cite{membrane}, immersed in a thermal environment at $1$ K, with initial
effective occupancy $n_{i}=10^{5}$ at $\omega _{m} $. Notice that $\kappa /\omega _{m}=5$
corresponds to the inefficient unresolved sideband optomechanical regime where, in the absence of
the atomic medium, cooling is slow and optimized around $\Delta _{c}\simeq \kappa /2$. For a
control field of Rabi frequency $\Omega =(2\pi )300$ MHz the effective cavity decay rate reduces to
$\kappa _{EIT}\simeq \kappa /10$, which puts us slightly in the resolved sideband regime. To
estimate the effective cooling of the scheme one has to fix the optomechanical effective coupling
$G$. For a membrane with an effective mass around $1$ ng in a 1 cm-long optical cavity one gets a
single-photon optomechanical coupling strength
of the order of $G_{0}\simeq (2\pi )200$ Hz. For a maximum $
\left\vert \left\langle a\right\rangle \right\vert \simeq 10^{3}$, to satisfy $g\left\vert \left\langle a\right\rangle \right\vert \ll \Omega $,
one would obtain $G\simeq (2\pi )200$ kHz. For these
parameters we numerically calculate the variance matrix from
Eqs.~(\ref{eq:c_3}-\ref{eq:b}) and show in Fig.~\ref{cooling}a the expected
optimization of cooling when the two photon resonance is matched to $\omega
_{m}$. The effective cavity window then completely includes the
Anti-Stokes sideband for efficient cooling, $%
\kappa _{EIT}\gg \Gamma _{O}=(2\pi )40$ kHz. Fixing $\delta =\omega _{m}$, Fig.~%
\ref{cooling}b shows the variation of the residual occupancy with the
control field Rabi frequency. As expected from Eq.~(\ref{eq:n_f}), the
occupancy decreases as $\kappa _{EIT}$ increases until the EIT window
becomes too large to resolve the sidebands. We show in Fig.~\ref{cooling}c a
comparison between cavity EIT cooling and standard optimized cavity cooling
(when no atoms are present) with fixed $\Delta _{c}\simeq \kappa /2$. The
obtainable temperature is about two orders of magnitude lower in the
EIT cooling case while the cooling rate is enhanced by a factor $\sim\kappa/\omega_m$.

We then check the validity of our RWA treatment indication of a strong
coupling regime by taking the example of a reversible state mapping of a
squeezed state. Considering an initial situation in which the atomic ground-state coherence has been prepared in a squeezed atomic state with squeezing
parameter $r=1$ and the mirror in an initial thermal state with average phonon
number $2$, we numerically integrate Eqs.~(\ref{eq:c_3}-\ref{eq:b}) and
calculate the time evolution of the atom and mirror Wigner functions. To
satisfy Eq.~(\ref{eq:strongcouplinginequality}), we take $\Omega=(2\pi )~100$ MHz and $G=(2\pi )$ $500$ kHz. The ratios of the coupling strength $\sqrt{C\gamma
_{E}\gamma _{O}}$ to the various decoherence rates for the chosen
illustration $(\gamma _{m}n_{i},\gamma _{O},\gamma _{E})$ are $%
(25,6.6~10^{4},5)$, showing the emergence of a strong coupling regime. Fig.~\ref{cooling}d shows
indeed the transfer of
the atomic squeezed state onto the mechanical resonator after a time $\pi /\sqrt{%
C\gamma _{E}\gamma _{O}}$. Taking $
\left\vert \left\langle a\right\rangle \right\vert \simeq 250$ (to still satisfy $g|\langle a
\rangle|\ll\Omega$) this would mean an optomechanical coupling rate $G_0\sim (2\pi)2 kHz$, a value
somewhat higher than that achieved with state-of-the-art SiN membranes~\cite{membrane}.

\begin{figure}[t]
\includegraphics[width=0.95\columnwidth]{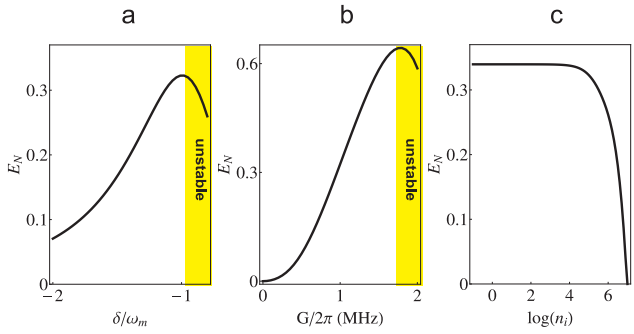}
\caption{\textit{Atom-mirror entanglement:} (a) Logarithmic negativity $E_{N}$ as a function of two-photon
detuning $\protect\delta $ for the parameters given in the text. (b) Variation of $E_{N}$ with $G$ for $%
\protect\delta =-\protect\omega _{m}$. c) Variation of $E_{N}$ with $n_{i}$,
showing the robustness of the entanglement with respect to temperature.}
\label{ent}
\end{figure}

\textit{Stokes sideband resonance: atom-mirror entanglement} We now turn to
the case where the cavity and the EIT medium are tuned to the Stokes
sideband. Assuming $\delta =\Delta_{c}=-\omega _{m}$ and neglecting again
off-resonant interactions, one gets a set of equations similar to Eqs. (\ref%
{RWA}), $\tilde{b}$ being replaced by $\tilde{b}^{\dagger }$. Eliminating the fast variables in the frame rotating at $-\omega
_{m}$, one can again deduce an effective Hamiltonian for the reduced atom-mirror
system which now takes the form of a parametric down-conversion process
\begin{equation}
H_{S}\simeq \sqrt{C\gamma _{E}\gamma _{O}}\left( \tilde{b}^{\dagger }\tilde{c%
}_{2}^{\dagger }+\tilde{b}\tilde{c}_{2}\right) ,
\end{equation}
known to generate bipartite entanglement from an initial bimodal separable
state~\cite{nielsen00}. To quantify this entanglement we calculate
the logarithmic negativity $E_{N}$~\cite{logneg} by numerically integrating Eqs.~(\ref{eq:c_3}-\ref{eq:b}). However,
a closer look into the RWA equations of motion shows that despite the fact
that the down-conversion process does lead to an entangled steady state, the
assumption of $\Delta_{c}=-\omega _{m}$ strongly limits the achievable
entanglement, owing to the occurrence of a parametric instability even for very
small values of $G$. To get around the limitation imposed by the parametric heating of the
membrane, one can use a far-detuned cavity such that $\left\vert \Delta
_{c}\right\vert \gg \kappa ,\omega _{m}$. In such a case higher values of $G
$ are allowed before the onset of parametric instability and considerable entanglement can in principle be generated, as illustrated
in Fig.~\ref{ent}. As an example, we consider the parameters used
for Fig.~\ref{cooling}, except for $N=10^{4}$ and $\Omega =(2\pi )1.2$ MHz and choose a
cavity detuning $\Delta _{c}=-12\kappa $. As shown in Fig.~\ref%
{ent}a, the entanglement is maximum around $\delta \simeq -\omega _{m}$,
close to the point where parametric instability occurs. The entanglement dependence on $G$
is shown in Fig.~\ref{ent}b and the expected increase with the cavity
driving is obtained. Under the condition $\delta \simeq -\omega _{m}$ and for $%
G=(2\pi )1$ MHz, Fig.~\ref{ent}c shows that the generated entanglement is
somehow quite robust with respect to the thermal environment of the
mechanical oscillator, as substantial entanglement is still present even at
temperatures of $\sim 20$ K. The point where the entanglement washes out can
also be analytically estimated by equaling the coupling rate to the thermal
decoherence rate. Adiabatic eliminations of the cavity field and atomic dipole result in an
effective atom-membrane coupling $\Omega g_{N}G/\sqrt{g_{N}^{4}+\gamma
^{2}\Delta _{c}^{2}}$. For the parameters considered, this coupling
equals the thermal decoherence rate at $n_{i}\simeq 6\times 10^{6}$, in agreement with Fig.~\ref{ent}c.

\textit{Conclusion and outlook} We have shown that a hybrid
optomechanical approach in dealing with quantum effects at the
mesoscale range defined by a mechanical resonator can be employed
for accessing regimes which would otherwise be inaccessible in the
bare optomechanical system. Strong coupling and entanglement in the
unresolved sideband regime of a cavity-membrane system can for
instance be engineered via the controllable atom-field EIT effect.
Conditioned by experimental progress in increasing the bare
optomechanical coupling, we envision a ion Coulomb crystal-mebrane
hybrid system where these effects can be verified. Moreover, when
instead of a intracavity membrane one considers the motion of the
Coulomb crystal as a whole as the mechanical degree of freedom,
cavity mediated motional-internal state coupling can be also
similarly shown.

\textit{Acknowlegments} MD and AD acknowledge support from the ESF
Euroquam \textquotedblright CMMC" and EU \textquotedblright PICC"
and \textquotedblright CCQED" projects. CG and HR acknowledge
support from the NanoSci-E+ Project \textquotedblright NOIs".

\end{document}